\documentclass[sigconf,nonacm]{acmart}

\AtBeginDocument{%
  }

\acmConference[Submission to UIST '26]{Make sure to enter the correct
  conference title from your rights confirmation email}{Nov 02--05,
  2026}{Detroit, MI}

\begin{document}

\title{TRAFA: Anticipating User Actions to Reduce Errors in Procedural Tasks with Predictive Feedback}

\author{Sassan Mokhtar}
\email{mokhtar@iai.uni-bonn.de}
\orcid{1234-5678-9012}
\affiliation{%
  \institution{University of Bonn}
  \city{Bonn}
  \country{Germany}
}

\author{Lars Doorenbos}
\email{doorenbos@iai.uni-bonn.de}
\affiliation{%
  \institution{University of Bonn}
  \city{Bonn}
  \country{Germany}
}
\affiliation{%
  \institution{Lamarr Institute for Machine Learning and Artificial Intelligence}
  \city{Dortmund}
  \country{Germany}
}

\author{Fatemeh Jabbari}
\email{jabbari@iai.uni-bonn.de}
\affiliation{%
  \institution{University of Bonn}
  \city{Bonn}
  \country{Germany}
}

\author{Marius Bock}
\email{bock@iai.uni-bonn.de}
\affiliation{%
  \institution{University of Bonn}
  \city{Bonn}
  \country{Germany}
}
\affiliation{%
  \institution{Lamarr Institute for Machine Learning and Artificial Intelligence}
  \city{Dortmund}
  \country{Germany}
}

\author{Dominik Bach}
\email{d.bach@uni-bonn.de}
\affiliation{%
  \institution{University of Bonn}
  \city{Bonn}
  \country{Germany}
}

\author{Juergen Gall}
\email{gall@iai.uni-bonn.de}
\affiliation{%
  \institution{University of Bonn}
  \city{Bonn}
  \country{Germany}
}
\affiliation{%
  \institution{Lamarr Institute for Machine Learning and Artificial Intelligence}
  \city{Dortmund}
  \country{Germany}
}
\renewcommand{\shortauthors}{Mokhtar et al.}

\newif\ifdraft
\drafttrue

\definecolor{orange}{rgb}{1,0.5,0}
\definecolor{gr}{rgb}{0,0.65,0}
\definecolor{mygray}{gray}{0.95}

\ifdraft
 \newcommand{\LD}[1]{{\color{blue}{\bf LD: #1}}}
 \newcommand{\ld}[1]{{\color{blue}#1}}
 \newcommand{\old}[1]{{\color{gr}#1}}
\else
 \renewcommand{\sout}[1]{}
\fi

\newcommand{\framework}{\textsc{\ld{TODO}}\xspace}
\definecolor{lgray}{gray}{0.9}
\newcommand{\real}{\mathbb{R}}
\newcommand{\x}{\mathbf{x}}
\newcommand{\z}{\mathbf{z}}
\newcommand{\y}{\mathbf{y}}
\newcommand{\haty}{\hat{\y}}
\newcommand{\w}{\mathbf{w}}
\renewcommand{\d}{\mathbf{d}}
\newcommand{\D}{\mathcal{D}}
\newcommand{\X}{\mathcal{X}}
\newcommand{\Z}{\mathcal{Z}}
\newcommand{\J}{\mathbf{J}}
\newcommand{\bZ}{\mathbf{Z}}
\newcommand{\M}{\mathcal{M}}
\newcommand{\I}{\mathcal{I}}
\newcommand{\jacobian}{\mathbf{J}}
\newcommand{\balpha}{\bm{\alpha}}
\newcommand{\pkde}{p_{\textrm{kde}}}
\newcommand{\psv}{p_{\balpha}}
\newcommand{\f}{\mathbf{f}}
\newcommand{\g}{\mathbf{g}}
\newcommand{\F}{\mathcal{F}}
\renewcommand{\a}{\mathbf{a}}

\newcommand{\MSCL}{{\bf{MSCL}}}
\newcommand{\PANDA}{{\bf{PANDA}}}
\newcommand{\MKD}{{\bf{MKD}}}
\newcommand{\SSD}{{\bf{SSD}}}
\newcommand{\DNtwo}{{\bf{DN2}}}
\newcommand{\MHRot}{{\bf{MHRot}}}
\newcommand{\DDV}{{\bf{DDV}}}
\newcommand{\IC}{{\bf{IC}}}
\newcommand{\HierAD}{{\bf{HierAD}}}
\newcommand{\Glow}{{\bf{Glow}}}
\newcommand{\MahaAD}{{\bf{MahaAD}}}
\newcommand{\CFlow}{{\bf{CFlow}}}
\newcommand{\NL}{{\bf{NL-Invs}}}
\newcommand{\DIF}{{\bf{DIF}}}

\newcommand{\uniclass}{\emph{uni-class}}
\newcommand{\uniano}{\emph{uni-ano}}
\newcommand{\unimed}{\emph{uni-med}}
\newcommand{\shiftlowres}{\emph{shift-low-res}}
\newcommand{\shifthighres}{\emph{shift-high-res}}

\newcommand{\thyroid}{\emph{thyroid}}
\newcommand{\bc}{\emph{breast cancer}}
\newcommand{\speech}{\emph{speech}}
\newcommand{\pg}{\emph{pen global}}
\newcommand{\shuttle}{\emph{shuttle}}
\newcommand{\kdd}{\emph{KDD99}}

\newcommand{\guood}{{\bf{General U-OOD}}}
\newcommand{\shd}{{\bf{Shallow U-OOD}}}

\newcommand{\xmark}{\ding{55}}%
\newcommand{\cmark}{\ding{51}}%



\begin{abstract}
Interactive assistance systems typically provide feedback after an action has been completed, supporting error recovery but not preventing the error itself. We present \textsc{TRAFA}, a real-time predictive feedback system for procedural tasks that intervenes before errors are committed.  \textsc{TRAFA} operationalizes predictive feedback through a Track-Forecast-Act framework that tracks hand and object state, forecasts user motion conditioned on scene context, and triggers feedback when a predicted action is likely to violate task constraints. We instantiate this pipeline in a sequential assembly setting and evaluate it through both technical benchmarking and a controlled user study against conventional reactive feedback. Our results show that predictive feedback improves task accuracy and efficiency while maintaining a comparable number of feedback events. These findings position feedback timing as a key dimension in system design and show how real-time anticipation can be integrated into interactive systems to prevent errors before they occur. 
\end{abstract}



\begin{teaserfigure}
 \centering
 \includegraphics[width=0.95\textwidth]{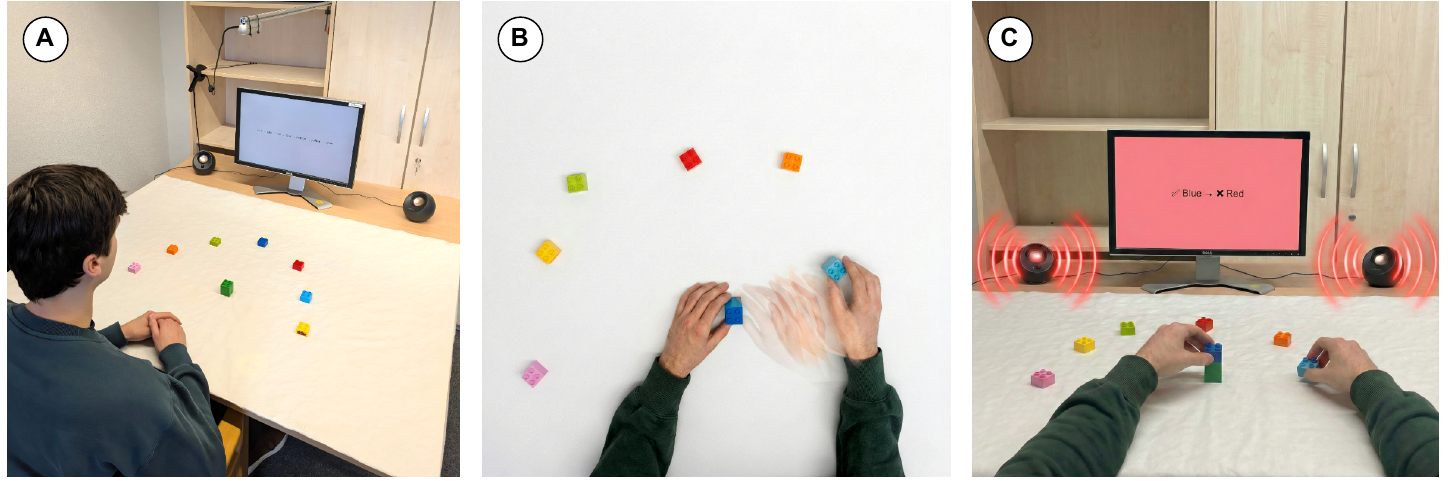}
 \caption{\textbf{Overview of the prototype of the predictive feedback system.} \textbf{(A)} A participant is tasked to assemble colored blocks in a specific order on a tabletop workspace, while the system observes hand motion and scene state. \textbf{(B)} The system forecasts near-future hand motion from recent observations, illustrated as fading hand trajectories, and combines these forecasts with scene context to anticipate the likely next placement. \textbf{(C)} When an imminent error, that is an incorrect placement, is anticipated, the system provides feedback through visual and auditory cues, enabling error prevention rather than post-hoc correction.}
 \label{fig:teaser_new}
\end{teaserfigure}

\maketitle

\section{Introduction}

Humans frequently make mistakes in everyday activities, and in many domains, these errors are costly to detect and recover from. For instance, over 50\% of surgical complications are directly caused by mistakes~\cite{suliburk2019analysis} and up to 90\% of defects in assembly line production systems stem from human errors~\cite{le2012novel}. While some errors can be corrected at minimal cost if detected in time, others are costly to undo or even irreversible, leading to substantial damages. Reducing human error is therefore a central challenge in domains where task safety, reliability, and efficiency matter.

Human-computer interaction (HCI) research has long addressed this challenge through feedback mechanisms that guide users toward correct task execution, with prior work exploring visual, auditory, and multimodal feedback across a wide range of procedural tasks. However, much of this work emphasizes feedback mechanisms that support post-hoc correction rather than anticipating errors before they occur~\cite{rossmy2023point}. Even systems that incorporate anticipatory sensing often evaluate prediction accuracy and early recognition rather than addressing how such predictions can be operationalized in a low-latency interactive loop that intervenes during ongoing action~\cite{li2025satori, zhang2022predicting, pawar2025earl}. Although some recent work has begun to model the timing of intelligent suggestions~\cite{yu2022optimizing}, the system design question of how to turn anticipation into timely and actionable feedback remains largely open.

\begin{figure*}[t]
    \centering
    \includegraphics[width=0.9\textwidth]{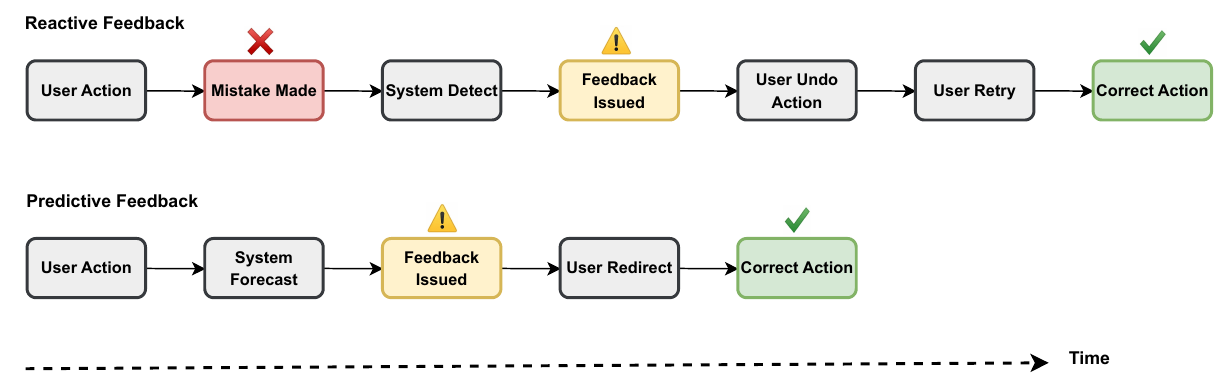}
    \caption{\textbf{Difference between reactive and predictive feedback.} Reactive feedback (top) intervenes after error completion, whereas predictive feedback (bottom) intervenes earlier based on anticipated action outcomes. In this way, predictive feedback redirects users before committing to errors.}
    \label{fig:feedback_flow}
\end{figure*}

In this work, we investigate \textit{predictive feedback}: feedback delivered while users can still redirect an ongoing action before making an error. Moving from reactive to predictive feedback requires more than detecting mistakes earlier; it requires a system that continuously tracks task-relevant states, anticipates near-future user motion, and decides when to intervene (Figure~\ref{fig:feedback_flow}). To address this challenge, we present  \textsc{TRAFA}, a real-time predictive feedback system that operationalizes this loop through a \textit{Track-Forecast-Act} paradigm: the system tracks hand and scene state, forecasts short-horizon hand motion, and triggers feedback when a predicted action is likely to violate task constraints (Figure~\ref{fig:teaser_new}). We instantiate this framework in a controlled sequential assembly task, where users reproduce a target color sequence by stacking interlocking blocks. In this setting, the system continuously predicts near-future hand trajectories and uses these forecasts together with scene state to assess whether an incorrect placement is likely to occur. By grounding intervention decisions in short-horizon motion prediction, the system can provide feedback while the action is still in progress, making predictive feedback practical for continuous procedural interaction.

We evaluate the proposed system in two ways. First, we benchmark the anticipation module against alternative designs to assess whether the system is accurate enough to support pre-commitment intervention. Second, we conduct a within-subject user study with 20 participants comparing predictive feedback against reactive feedback and a no-feedback baseline across auditory, visual, and multimodal conditions. Across conditions, predictive feedback significantly improves task accuracy and efficiency relative to reactive feedback, while maintaining a comparable number of feedback events. These findings show that the benefit of the proposed system comes not from issuing more feedback, but from issuing it early enough to prevent an incorrect action from being completed. Together, the results demonstrate how real-time anticipation can be operationalized in an interactive system to support error prevention in procedural tasks.

Our main contributions are as follows:
\begin{enumerate}
    \item A \textit{Track-Forecast-Act} system for predictive feedback that couples scene tracking, short-horizon motion forecasting, and intervention logic in real time.
    \item A scene-aware anticipation and intervention method that uses hand motion forecasts together with the task state to trigger feedback before an incorrect placement is made.
    \item An evaluation combining technical benchmarking and a controlled user study, showing that predictive feedback improves task accuracy and efficiency without increasing feedback frequency.
\end{enumerate}

\section{Related Work}
Feedback is a foundational concept in human–computer interaction, supporting error awareness, learning, and task regulation. Classic HCI frameworks emphasize feedback as a way to keep users informed about the system state and results of their actions, helping them understand what has occurred and to recognize and recover from errors~\cite{norman2013design, nielsen1994usability}. Prior work has demonstrated that feedback design strongly influences how users interpret system behavior and regulate their actions~\cite{howie2000human}.

A large body of research has explored feedback content and modality, most commonly visual, auditory, and tactile feedback. While tactile systems have been shown to be able to steer human movements \cite{bial2011enhancing}, this work specifically focuses on audio and visual feedback and their combination. Visual and auditory feedback have been widely studied as a reactive mechanism, signaling errors or confirming correctness after an action has been completed in sequential and procedural tasks \cite{rauterberg1994positive, funkUsingInSituProjection2015}. Prior studies show that modalities can improve learnability, usability, and task performance~\cite{poelman2010effect, funkHapticAuditoryVisual2016}. At the same time, visual feedback can be distracting in cognitively demanding tasks~\cite{yatani2012investigating}, motivating research on auditory and multimodal alternatives. Early works on non-speech audio showed that simple auditory cues can effectively convey errors and status without requiring visual attention~\cite{hereford1994non}, while later work examined how different modalities can play complementary roles in interaction~\cite{frauenberger2009auditory, freeman2017multimodal, schaffertReviewRelationshipSound2019, yinEvaluatingMultimodalFeedback2019}. 

Beyond feedback modality, HCI research has long emphasized the importance of anticipation in interaction. Early work argued that systems should help users understand the expected outcomes of their actions rather than relying solely on feedback after action completion~\cite{djajadiningrat2002but}. This perspective motivated feedforward design strategies, in which expected action outcomes are revealed prior to commitment to support planning and decision-making~\cite{vermeulen2013crossing}. Feedforward mechanisms, however, are typically defined in advance by the interface and present static previews of possible outcomes, independent of the user’s ongoing motion or moment-to-moment action execution. In parallel, research in motor control shows that the timing of information, such as delayed or anticipatory feedback, can influence how people coordinate actions and engage in anticipatory behavior~\cite{tramper2013feedforward, roman2019delayed}. Together, these perspectives suggest that not only the content of feedback, but also its timing relative to ongoing action, can shape user behavior.

At the same time, prediction of human motion and intent has been widely studied in interactive and collaborative systems, particularly in robotics and assistive domains. Prior work used motion forecasting, and in some cases intent inference, to enable smoother handovers, safer collaboration, and more efficient coordination by allowing systems to adapt their behavior in anticipation of human actions~\cite{gomez2025enhancing, laplaza2025enhancing, noppeney2024human}. Related research has also leveraged human motion prediction to provide anticipatory warning or corrective feedback in ergonomics and posture training systems~\cite{billast2024physical}. Other research has explored real-time intention recognition using multimodal signals to support assistive or cooperative behavior~\cite{chiossi2025designing, chowdhury2025real}, or to enable anticipatory intervention in domains such as autonomous driving~\cite{johns2017looking}. In most of these systems, prediction is primarily used to guide system behavior, such as motion planning or safety constraints, rather than to determine when evaluative feedback should be delivered to the user.

Although prior work offers extensive guidance on feedback content, modality, and adaptivity in procedural tasks \cite{funkStopHelpingMe2015, precupRecognisingWorkerIntentions2023}, feedback timing is treated as an implicit consequence of system state changes rather than an explicit interaction design variable. Recent work on context-aware procedural assistants further motivates this area: prior work has explores smartwatch-based task support through step tracking, context-aware assistance, and proactive intervention, while later work scales such assistants through demonstration-based learning and mixed-initiative dialogue~\cite{arakawa2023prism, arakawa2024prism, arakawa2024prism1, arakawa2025scaling}. 
Related work on work interruptions~\cite{puranik2020pardon}, robotic decision support~\cite{natarajan2024trust} and coactive design \cite{johnson2014coactive} emphasizes that supportive systems should preserve observability, predictability, and directability in joint activity.
To our knowledge, this work presents the first predictive feedback system forecasting hand motion to control the timing of feedback delivered to a user during an ongoing task. This framing positions feedback timing as a central design concern and motivates an empirical investigation of how earlier intervention affects performance and behavior in a procedural task.

\section{Method}

\begin{figure*}[t]
  \centering
  \includegraphics[width=1.0\textwidth]{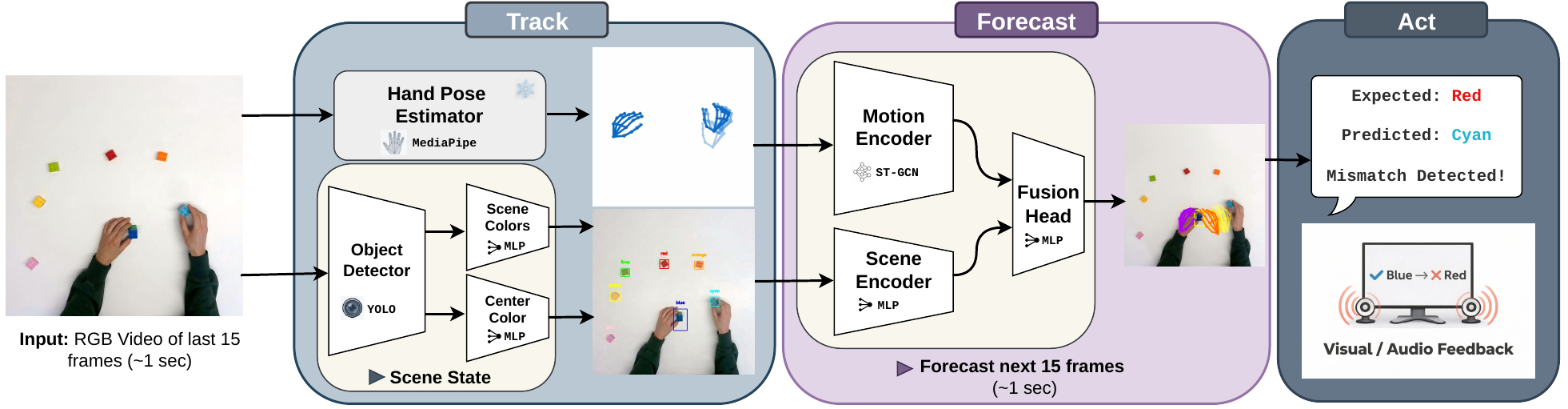}
  \caption{\textbf{Overview of \textsc{TRAFA}, our Track-Forecast-Act system.} The system takes as input the last 15 observed RGB frames. The \textbf{Track} module estimates hand pose and scene state from the detected objects. The \textbf{Forecast} module predicts the hand pose for the next 15 frames using a scene-aware forecasting model composed of a motion encoder, scene encoder, and fusion head. The \textbf{Act} module uses the predicted motion and current task state to anticipate whether the next placement will violate task instructions, and issues visual and/or auditory feedback before the incorrect action is done.}
  \label{fig:tfa_overview}
\end{figure*}

We consider procedural tasks in which users manipulate physical objects given task instructions, and where an incorrect action becomes costly once it is completed. In such settings, conventional feedback is often reactive: the system detects an error only after the user has already committed the action, requiring recovery through undoing or correction. Our goal is to enable \textit{predictive feedback}, in which the system intervenes while the action is still in progress and the user can still change course.

Realizing predictive feedback requires solving three coupled problems in real time. First, the system must maintain a task-relevant state, including the current configuration of objects and the user’s recent interactions with them. Second, it must anticipate the user’s near-future motion early enough for intervention to remain actionable. Third, it must translate predicted motion and task state into feedback decisions that are specific enough to prevent errors without over-triggering unnecessary alerts. We address this challenge in the context of a sequential assembly task, but design the system around a more general requirement shared by many procedural interactions: anticipating likely action outcomes before commitment.

To implement predictive feedback, we introduce \textsc{TRAFA}, a \textit{Track-Forecast-Act} system (Figure~\ref{fig:tfa_overview}). The system takes as input a short RGB video sequence and produces feedback when the predicted next action is likely to violate task constraints. The pipeline consists of three stages: \textit{Track}, which estimates task-relevant hand and scene state; \textit{Forecast}, which predicts short-horizon future hand motion conditioned on scene context; and \textit{Act}, which converts predicted motion and task state into feedback decisions. 

\subsection{Track: Perception and Task-State Estimation}
\label{sec:track}
The \textit{Track} stage obtains and maintains the task-relevant state required for predictive intervention. In our setup, which will be described in more detail in Sec.~\ref{sec:task_instantiation}, observations come from an overhead RGB camera observing a tabletop sequential assembly task, where users are tasked to stack seven distinct Duplo blocks in a given order. From this video stream, the system estimates both the current scene configuration and the user’s recent interaction history. This includes detecting the available blocks in the workspace, inferring their colors and positions, identifying the current top block in the assembly region, and tracking which block the user most recently interacted with.

To maintain this state, we use a lightweight multi-stage vision pipeline. First, a YOLOv8n detector~\cite{ultralytics_yolov8, redmon2016you} identifies all Duplo blocks in the workspace. The detector was trained on a custom dataset generated through a semi-automated annotation pipeline: SAM2~\cite{ravi2025sam} was used to produce initial segmentation masks on recorded task trajectories, which were then manually verified and corrected. The resulting dataset comprises approximately $900$ annotated frames. Following detection, block colors are inferred using  3-layer MLPs applied to masked image regions. We use two specialized MLPs: $(1)$ a scene-level color classifier that assigns a color label to each detected block in the workspace, and $(2)$ a center-piece classifier that determines the color of the top block on the base plate. Separating these two predictions improves robustness when the center region is partially occluded by the user’s hands.

In parallel, hand pose is extracted from the RGB stream using MediaPipe~\cite{lugaresi2019mediapipe}. Recent hand-object interactions are inferred from spatial overlap between the detected block bounding boxes and the fingertips' location. The estimated hand trajectories are also used to provide the motion history required for forecasting. Detected blocks are assigned discrete symbolic states (e.g., \textit{resting}, \textit{candidate}, \textit{placed}) based on their spatial location and interaction history. This scene representation decouples downstream forecasting and intervention logic from raw detector outputs and enables low-latency reasoning about likely placement outcomes.

\subsection{Forecast: Scene-Aware Motion Anticipation}
\label{sec:forecast}

\begin{figure*}[t]
    \centering
    \includegraphics[width=\linewidth]{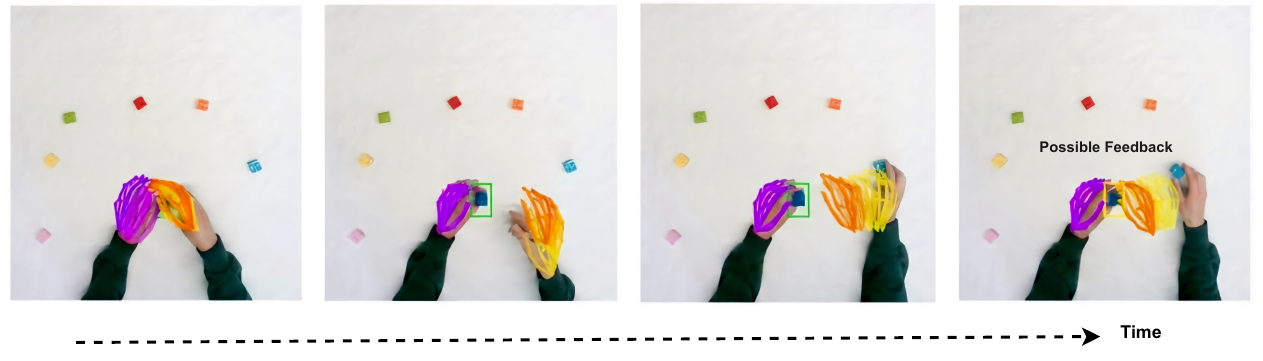}
    \caption{\textbf{Example of predictive intervention from forecast hand motion.}  From recent hand observations, the system forecasts near-future hand trajectories (colored curves). If the forecast motion is expected to enter the assembly region and the most recently interacted block does not match the next required block, the system anticipates an incorrect placement and issues feedback before the action is completed.}
    \label{fig:motion_forecasting}
\end{figure*}

Given the tracked scene state and recent hand motion, the \textit{Forecast} stage forecasts the user’s hand trajectory as illustrated in Figure~\ref{fig:motion_forecasting}. We formulate this as a conditional motion forecasting problem: rather than predicting motion in isolation, the model conditions future hand motion on the spatial configuration of objects, since actions in assembly tasks are strongly constrained by the locations of candidate blocks and the assembly region.

We implement this stage using a scene-aware Spatio-Temporal Graph Convolutional Network (ST-GCN)~\cite{yan2018spatial}. The model takes as input the last $15$ observed hand-pose frames (approximately one second) and predicts the next $15$ frames. Hand motion is represented as a graph over $42$ keypoints (two hands), with per-keypoint features given by 2D image coordinates and instantaneous velocities. A motion encoder based on stacked ST-GCN blocks extracts a compact representation of recent hand dynamics. In parallel, the bounding boxes of the detected blocks are encoded by a lightweight MLP to produce a scene embedding summarizing the spatial layout of the workspace. The motion and scene embeddings are fused and decoded to predict future hand poses over the forecasting horizon. The full forecaster contains approximately $210{,}000$ trainable parameters.

The one-second prediction horizon was chosen to preserve actionability: it is long enough to be perceived and acted upon during an ongoing movement, given typical human visual-motor response latencies of $200-300\  ms$~\cite{card2018psychology}, but short enough to avoid the uncertainty associated with longer-range motion prediction. To support reliable intervention, the model is trained with the objective
\begin{equation*}
\mathcal{L}=\mathcal{L}_{tip}+\beta\mathcal{L}_{pose}+\lambda\mathcal{L}_{smooth}
\end{equation*}
where $\mathcal{L}_{tip}$ is a temporally weighted mean squared error on fingertip positions, $\mathcal{L}_{pose}$ is the mean squared error over all hand keypoints, and $\mathcal{L}_{smooth}$ penalizes velocity inconsistency between predicted and ground-truth trajectories. This design emphasizes fingertip accuracy, which is most relevant for assembly tasks. We use $\beta=0.1$ and $\lambda=0.05$. 

\subsection{Act: Intervention Logic and Feedback Policy}
\label{sec:act}

The \textit{Act} stage converts predicted motion and current task state into intervention decisions. 
A predictive alert is triggered if the forecast hand trajectory is predicted to enter the assembly region $\mathcal{R}$, where the assembly happens, within the next $15$ frames and the block most recently interacted with does not match the next required color in the sequence. In this way, the system uses anticipated motion and current task state to infer that an incorrect placement is imminent, allowing intervention during the ongoing action rather than after the error has already been made.

\begin{figure*}
  \centering
  \includegraphics[width=0.95\textwidth]{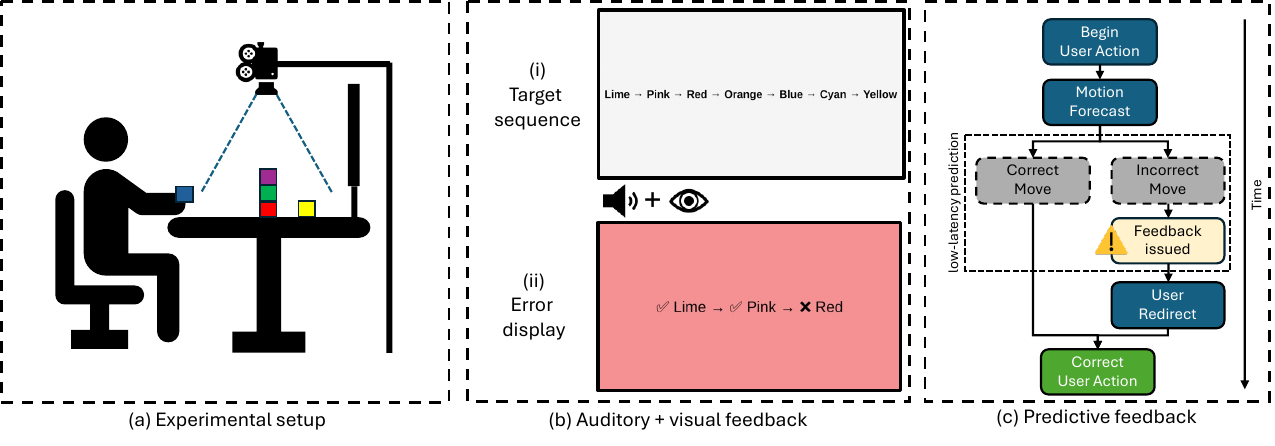}
 \caption{\textbf{Experimental setup and feedback paradigm.} \textbf{(a)} Participants performed a tabletop sequential assembly task while an overhead camera captured the workspace. \textbf{(b)} The monitor displayed (i) the target color sequence at the beginning of each trial and (ii) the visual error display; auditory cues could be presented together with the visual display. \textbf{(c)} In the predictive condition, the system forecasted the user’s ongoing motion and issued feedback before an incorrect placement was completed, allowing the user to redirect the action toward the correct move.}
 \label{fig:setup}
\end{figure*}

The feedback implementation itself is modular: the system can render visual feedback, auditory feedback, or both. Visual feedback is presented on the monitor by flashing the screen red and displaying the expected color together with the incorrect one as illustrated in Figure~\ref{fig:setup}. Auditory feedback is delivered through speakers using a warning tone followed by a spoken color cue. The experimental comparison of predictive, reactive, and no-feedback conditions across these output modalities is described in Section~\ref{sec:user_study_protocol}.

\begin{figure}[t]
\includegraphics[width=\columnwidth]{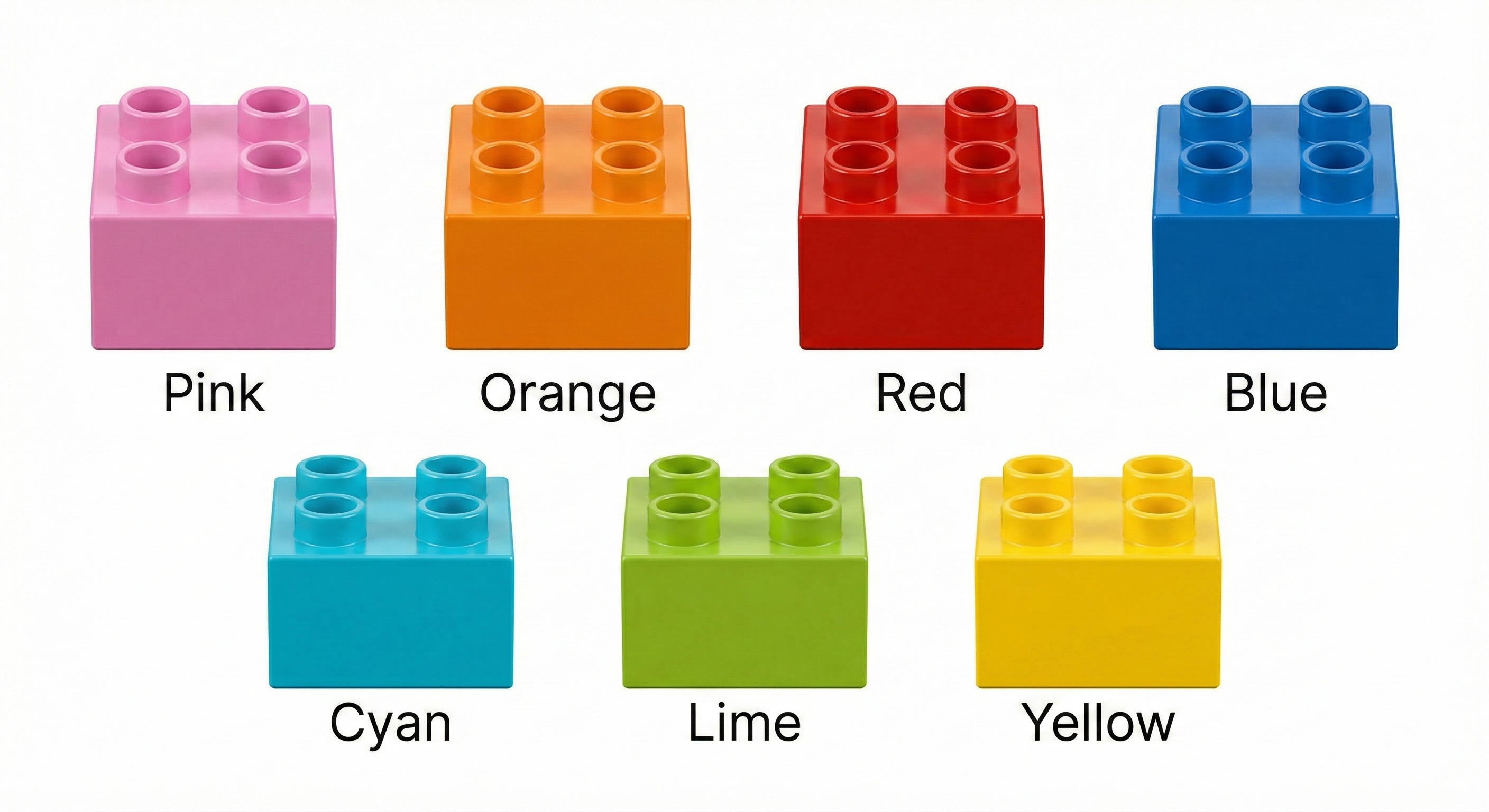}
\caption{Seven distinct colored Duplo building blocks used during the experiment.}
\label{fig:duploblocks}
\end{figure}

\subsection{Task Instantiation}
\label{sec:task_instantiation}

We instantiate the proposed system in a tabletop sequential assembly task as illustrated in Figure~\ref{fig:setup}. The workspace consists of seven distinct Duplo\footnote{Duplo blocks are large interlocking plastic building blocks designed for young children, \url{www.lego.com/en-us/themes/duplo/about}} blocks (Red, Orange, Yellow, Blue, Lime, Cyan, Pink) arranged around a fixed green base plate on a matte white surface (Figures~\ref{fig:motion_forecasting} and \ref{fig:duploblocks}). Before each trial, the seven blocks were placed in fixed spatial locations around the base plate, while the color assigned to each location was randomized. Participants were free to grasp, move, and reorganize pieces during the task.

At the beginning of each trial, a three-second countdown preceded the presentation of a target sequence of the seven colors on a $24$-inch monitor positioned $90$\,cm in front of the participant as illustrated in Figure~\ref{fig:setup}. The sequence was shown as written color names for one second and then removed. Participants were then asked to reproduce the target sequence by stacking the corresponding blocks onto the central base plate in the correct order.

We define errors in the following way: let $\mathcal{R}$ denote the assembly region, defined as a fixed rectangular area of $95\times60$ pixels centered on the base plate. An error occurs when a block is placed in $\mathcal{R}$ and its color does not match the next required color in the target sequence. Touching, grasping, or moving an incorrect block without placing it on the base plate was not considered an error. This conservative definition is typical in assembly tasks~\cite{klages2024human}, since it preserves exploratory behavior and allows participants to inspect, rearrange, and reconsider pieces without triggering feedback prematurely.

This task provides a controlled testbed for predictive feedback for three reasons. First, it has a clearly defined symbolic task state, since correctness depends on the next required block in the sequence. Second, errors become identifiable only at placement time, while still leaving a short pre-commitment window during the ongoing reach-to-place motion. Third, the task permits exploratory behavior such as touching or rearranging blocks before placement, making it suitable for studying when intervention should occur without over-triggering feedback.

All tracking, forecasting, and feedback components were implemented as ROS2 nodes~\cite{macenski2022robot} and executed on a workstation equipped with an NVIDIA TITAN RTX GPU (24\,GB VRAM). The workspace was captured using an overhead Intel RealSense D455 camera mounted perpendicular to the table surface. The camera streamed RGB video at a resolution of $1280\times720$ pixels and $15$ FPS. For processing, frames were cropped to a $720\times720$ region containing only the workspace.
Hand pose was extracted from the RGB stream using MediaPipe~\cite{lugaresi2019mediapipe} at approximately 15\,Hz, which served as the control frequency for forecasting and intervention. Across all components, the full system achieved a mean effective processing rate of $15.02$ FPS (SD$=0.58$), corresponding to an average processing cycle of approximately $67$\,ms per frame. This runtime is below the typical human visual--motor response latency~\cite{card2018psychology}, allowing the system to issue feedback while the user’s action is still in progress.

\subsection{User Study Protocol}
\label{sec:user_study_protocol}

To evaluate the interaction effects of predictive feedback, we conducted a within-subject user study with 20 participants (15 male, 5 female; age 23--38 years, $M=29$, $SD=4.3$), recruited at university campus.

None of the participants was affiliated with any of the authors or had used the system previously. All participants provided their informed consent. Participants took part individually. Each session lasted approximately 15 to 20 minutes and was conducted in the same laboratory room, with the door closed and the overhead lights turned on throughout the experiment. Two researchers were present throughout each session to provide instructions, operate the experimental scripts, and set up the blocks in the appropriate randomized order at the start of each trial. 
To limit the effect of external interventions during the experiment, researchers refrained from talking to participants, yet allowed them to ask questions if something remained unclear.

Participants received printed task instructions before the experiment. To avoid biasing participants toward a specific feedback modality or particular movement strategies through demonstration, no practice trials or demonstration videos were provided to participants prior to their first session. Each participant completed two experimental rounds to reduce noise from single-trial variability, with results then being averaged per participant for analysis. In each round, participants performed the task once under each of the seven feedback conditions, resulting in a total of $2 \times 7 \times 20 = 280$ trials. To eliminate measured differences between each feedback trial being the result of participants getting accustomed to the task, a new target color sequence was randomly generated for each individual trial, and the order of the seven feedback conditions was randomized within each round. Short breaks were allowed to be taken by participants at any point in the experiment to reduce fatigue.

Across all conditions, the task, sensing pipeline, and feedback representations were identical; only the timing and modality of feedback differed:

\begin{itemize}
    \item \textbf{No Feedback:} no feedback was provided during the task.
    \item \textbf{Reactive Audio:} auditory feedback was provided only after the system detected an incorrect placement.
    \item \textbf{Reactive Visual:} visual feedback was provided only after the system detected an incorrect placement.
    \item \textbf{Reactive Audio+Visual:} both auditory and visual feedback were provided only after the system detected an incorrect placement.
    \item \textbf{Predictive Audio:} auditory feedback was provided when the system predicted an imminent incorrect placement.
    \item \textbf{Predictive Visual:} visual feedback was provided when the system predicted an imminent incorrect placement.
    \item \textbf{Predictive Audio+Visual:} both auditory and visual feedback were provided when the system predicted an imminent incorrect placement.
\end{itemize}

After completing both rounds, participants filled out a brief post-study questionnaire. In addition to demographic information, the questionnaire asked participants to indicate their preferred feedback condition, rank the feedback modalities (audio, visual, and audio+visual), and optionally provide free-text comments explaining their preferences. These responses were used to complement the quantitative analysis.



\section{Results}

To evaluate the effectiveness of predictive feedback relative to other conditions, we measure task performance using a set of complementary metrics:

\begin{itemize}
    \item \textbf{Success Rate} measures whether a trial was considered successful (binary). A trial is considered successful if and only if the sequence of blocks placed on the base plate exactly matches the target sequence. Any trial containing at least one incorrect placement was counted as a failure, even if the incorrect block was later removed or corrected. 
    \item \textbf{Edit Distance} captures stepwise differences in performance beyond binary success using the edit distance. Specifically, we computed the minimum number of insertions, deletions, or substitutions required to transform the participant’s placed sequence into the target sequence. 
    \item \textbf{Feedback Frequency} assesses the cost of assistance, measured by the average number of feedback events per trial. 
    \item \textbf{Efficiency} measures the number of correctly placed blocks per minute. 
    \item \textbf{Total time} is the time taken for the trial.
\end{itemize}

As several metrics are bounded or non-normally distributed (e.g., binary success and discrete edit distance), we used non-parametric tests for all within-subject comparisons. Specifically, we applied two-sided Wilcoxon signed-rank tests to participant-averaged outcomes when comparing predictive, reactive, and no-feedback conditions. 
To control family-wise error across these four related comparisons, we applied the Holm--Bonferroni procedure. In addition to $p$-values, we report Wilcoxon effect sizes as $r = Z / \sqrt{N}$, where $N=20$ participants. 
We complement the metrics with participant-level distribution plots showing medians, interquartile ranges, and individual data points.

\subsection{Main Results}
\textbf{Predictive Feedback Improves Accuracy and Efficiency.}
Table~\ref{tab:res_sum} shows that predictive feedback substantially improved task accuracy compared to both reactive feedback and no feedback. When aggregated across modalities, predictive feedback increased the average task success rate from $15.0\%$ in reactive and $12.5\%$ in no-feedback conditions to $65.0\%$. Similarly, the edit distance decreased from an average of $2.06$ and $2.70$ errors per trial to $0.74$ with predictive feedback. The Wilcoxon signed-rank tests confirmed that predictive feedback significantly outperformed reactive feedback in both success rate ($p = 0.0001$) and edit distance ($p = 0.0001$).

Furthermore, predictive feedback led to a significant increase in efficiency, from $7.7$ correct blocks per minute to $8.7$ ($p = 0.012$), indicating faster progress through the task while maintaining correctness. Predictive feedback did not lead to a significant reduction in overall task completion time. This is expected given the low cost of errors in the task, where incorrect blocks could easily be removed and corrected without much penalty. Instead, the gains in efficiency arise from fewer incorrect placements and corrections.
Together, these findings demonstrate that predictive feedback improves both absolute and relative accuracy while increasing task efficiency, showing its effectiveness for preventing errors in sequential manual assembly tasks.

Notably, reactive feedback resulted in success rates comparable to the no-feedback conditions, while yielding lower edit distances. This reflects the timing constraints of reactive feedback. Since it is delivered only after an incorrect placement is completed, it cannot prevent task failure in a strictly ordered sequence. However, it can still reduce the number of subsequent errors once a mistake has occurred. No-feedback trials produced the highest edit distances overall, indicating that feedback in general supports reliable task completion.

\renewcommand{\arraystretch}{1.2}
\begin{table*}[t]
\centering
\caption{\textbf{Evaluating the effectiveness of predictive feedback in our user study.} We report five performance metrics for the seven different feedback conditions (mean $\pm$ SD across participants, $N=20$) and their averages, with the best numbers shown in bold.
Predictive feedback significantly improves the absolute task success rate and decreases edit distance without increasing feedback frequency. Aggregated rows report participant-level averages pooled across feedback modalities within each feedback timing condition.} 
\label{tab:res_sum}
\begin{tabular}{lccccc}
\toprule
\textbf{}                    & \textbf{\begin{tabular}[c]{@{}c@{}}Success \\ Rate\end{tabular}} $\uparrow$ & \textbf{\begin{tabular}[c]{@{}c@{}}Edit \\ Distance\end{tabular}} $\downarrow$    & \textbf{\begin{tabular}[c]{@{}c@{}}Feedback \\ Frequency\end{tabular}} $\downarrow$ & \textbf{\begin{tabular}[c]{@{}c@{}}Efficiency\\ (\# correct/min)\end{tabular}} $\uparrow$ & \textbf{\begin{tabular}[c]{@{}c@{}}Total time \\ (in sec)\end{tabular}} $\downarrow$ \\ 
\midrule
\textbf{No Feedback}              & 12.5\%              & 2.70 (±0.98)                     & -            & 7.3 (±3.4)                 & 33.1 (±10.7)         \\ 
\midrule
\textbf{Reactive Audio}           & 17.5\%              & 1.98 (±1.02)                     & 2.3 (±1.5)   & 7.6 (±3.2)                 & 46.0 (±7.1)          \\
\textbf{Reactive Visual}        & 10.0\%                 & 2.30 (±1.20)                      & 2.6 (±1.5)   & 7.3 (±3.2)                 & 49.3 (±11.8)          \\
\textbf{Reactive Audio+Visual}  & 17.5\%                  & 1.90 (±1.01)                      & \textbf{2.1 (±1.0)}   & 8.3 (±2.5)                 & 46.1 (±9.9)          \\ 
\textbf{Reactive (aggregated)}   & 15.0\%   & 2.06 (±1.08)  & 2.3 (±1.3) & 7.7 (±3.0) & 47.1 (±9.6) \\ 
\midrule
\textbf{Predictive Audio}       & \textbf{70.0\%}     & 0.75 (±1.19)             & 2.5 (±1.5)        & \textbf{9.1 (±2.7)}   & 44.4 (±10.3) \\
\textbf{Predictive Visual}       & 60.0\%         & 0.93 (±1.28)                      & 2.7 (±1.6)        & 8.0 (±2.0)             & 49.7 (±8.9)          \\
\textbf{Predictive Audio+Visual} & 65.0\%     & \textbf{0.53 (±0.62)}             & 2.3 (±1.3)        & 8.9 (±2.4)             & 46.7 (±9.0)          \\ 
\textbf{Predictive (aggregated)} & 65.0\%  & 0.74 (±1.03)  & 2.5 (±1.5) & 8.7 (±2.4) & 46.9 (±9.4) \\ 
\bottomrule
\end{tabular}
\end{table*}

\begin{figure*}[t]
  \centering
  \includegraphics[width=0.95\textwidth]{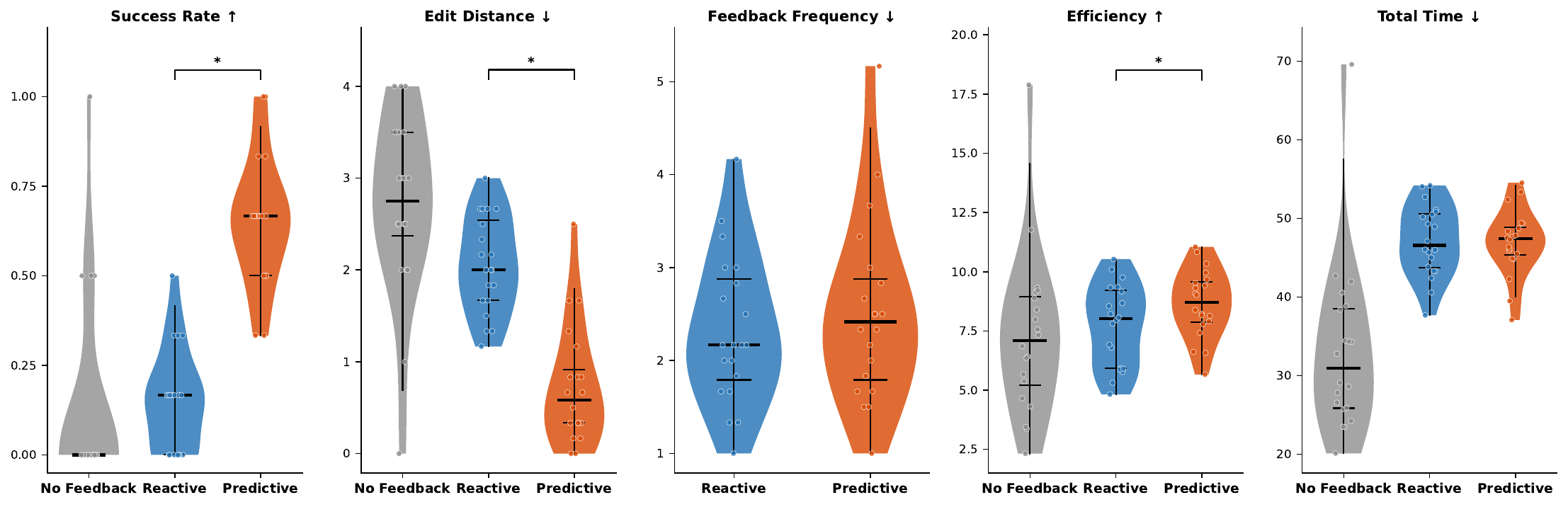}
 \caption{\textbf{Distribution of participant-level metrics across conditions ($N=20$).} Each violin shows the kernel density estimate; horizontal bars indicate the median and interquartile range; dots are individual participant values. Stars denote pairwise comparisons between Reactive and Predictive conditions that were statistically significant (Wilcoxon signed-rank, $p<0.05$). 
 }
 \label{fig:viollin}
\end{figure*}

\textbf{Predictive Feedback Does Not Increase Feedback Frequency.}
A central design goal of predictive feedback is to improve task performance without increasing the amount of feedback delivered, as excessive or frequent feedback can disrupt user attention and increase cognitive load~\cite{adamczyk2004if}. To assess this, we compared the number of feedback events across reactive and predictive feedback conditions. As shown in Table~\ref{tab:res_sum}, the average number of feedback events did not differ significantly between predictive and reactive feedback. Wilcoxon signed-rank tests revealed no significant effect of feedback type on feedback frequency ($p = 0.57$). Thus, the accuracy and efficiency gains observed with predictive feedback were not achieved by issuing more feedback. Instead, predictive feedback improved performance by changing \emph{when} feedback was delivered, intervening earlier in the action sequence before incorrect placements were completed. All reported main effects remained significant under family-wise correction.

Figure~\ref{fig:viollin} visualizes the participant-level distributions underlying all comparisons. The distributional view is consistent with the non-parametric analysis: predictive feedback increased success rate and reduced edit distance relative to reactive feedback, while feedback frequency remained comparable. After Holm correction across the four primary outcomes, predictive feedback remained significantly better in success rate ($p_{\mathrm{corr}}=0.0003$, $r=0.88$), edit distance ($p_{\mathrm{corr}}=0.0003$, $r=0.88$), and efficiency ($p_{\mathrm{corr}}=0.0240$, $r=0.56$), whereas feedback frequency did not differ significantly ($p_{\mathrm{corr}}=0.5730$, $r=0.13$). Participant-level paired-difference plots for the four primary metrics are provided in Appendix~\ref{app:1} to visualize the within-subject effect patterns underlying these aggregate results.

\subsection{Modality Analysis}
We next examined whether feedback modality (Audio, Visual, or Audio+Visual) influenced task performance or feedback cost for predictive feedback. 
Wilcoxon signed-rank tests revealed no statistically significant differences between audio and visual feedback for success rate ($p = 0.24$), edit distance ($p = 0.25$), efficiency ($p = 0.08$), or feedback count ($p = 0.29$). Similarly, no significant differences were observed between Audio-only and Audio+Visual feedback for success rate ($p = 0.35$), edit distance ($p = 0.38$), efficiency ($p = 0.26$), or feedback count ($p = 0.51$).\par
Although none of these comparisons reached statistical significance, audio feedback showed slightly higher success rates and efficiency than visual-only feedback (Table~\ref{tab:res_sum}). These trends suggest a possible advantage of auditory cues for maintaining task flow, but given the absence of significant effects, modality differences should be interpreted cautiously.

\subsection{User Preferences}

\begin{figure}[t]
    \centering
    \includegraphics[width=\linewidth]{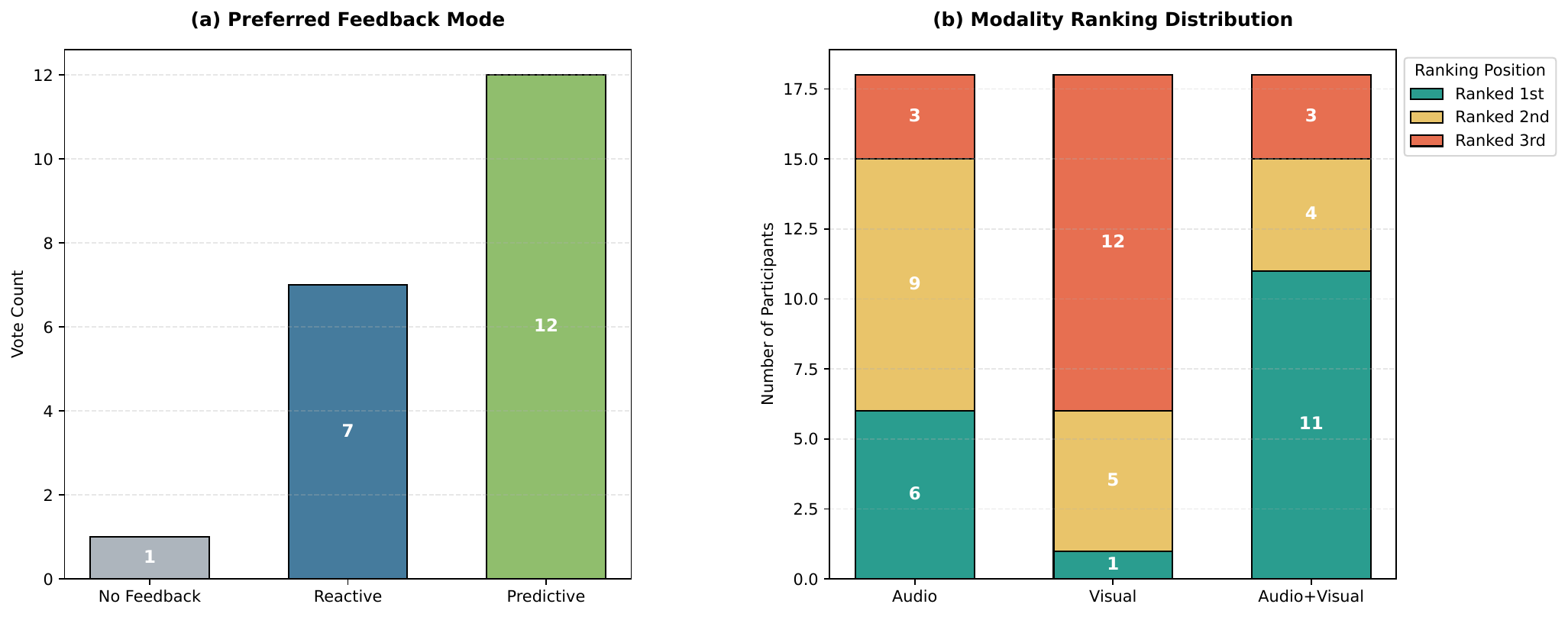}
    \caption{\textbf{Participant preferences from the post-study questionnaire.} (a) Preferred feedback mode across all conditions. Predictive feedback was preferred by most participants. (b) Ranking distribution for audio, visual, and audio+visual feedback modalities. Although audio was generally preferred over visual feedback, the combined condition received the strongest overall preference.}
    \label{fig:questionnaire}
\end{figure}

A post-study questionnaire completed by all participants after exposure to all feedback conditions provided additional insight into their subjective experiences. A majority of participants reported a preference for predictive feedback over reactive feedback (12 vs. 7), with one participant preferring the no-feedback condition (see Figure~\ref{fig:questionnaire}). Participants’ explanations emphasized differences in how feedback supported ongoing action. While most responses favored predictive feedback, some concerns were raised: one participant described predictive audio feedback as \textit{“sometimes too aggressive”} though this effect was mitigated when combined with visual cues. Overall, the stronger preference for predictive feedback suggests that participants valued assistance that helped them avoid errors in advance and did not perceive such intervention as intrusive.

Across feedback modalities, many participants expressed a preference for auditory over visual feedback, describing audio cues as \textit{“fast”} and \textit{“more direct”}, allowing them to maintain visual attention on the workspace. As one participant remarked, \textit{“The sound let me know immediately without having to look away from the blocks”}. In contrast, visual feedback alone was often perceived as requiring users to actively shift attention and \textit{“actively look for feedback”}. Nevertheless, visual feedback was valued for its ability to convey contextual information, such as task state and progress. Consequently, the combination of auditory and visual feedback was most frequently preferred overall. Participants characterized audio feedback as an immediate warning signal and visual feedback as supporting error interpretation and task understanding, as illustrated by one participant’s comment: \textit{“The sound caught my attention, and the visual feedback helped me see what went wrong.”}. These findings align with prior work demonstrating that different feedback modalities provide distinct and complementary strengths, with visual feedback particularly well suited for identifying required components and illustrating assembly actions~\cite{funkHapticAuditoryVisual2016}.

\subsection{Forecasting Module Ablation}
\label{sec:eval}
We justify our design of the forecasting module and evaluate whether it is more reliable to use for predictive feedback compared to other baselines. Table~\ref{tab:bench} compares four approaches: a linear extrapolation baseline, an immediate-placement baseline, a vanilla ST-GCN, and our scene-aware forecasting model. 
The linear extrapolation baseline that does not use learning: it extrapolates future hand motion directly from the recent pose history. The immediate-placement baseline assumes that a block will be placed next at the center as soon as the system detects contact between the hand and a block's bounding box. The vanilla ST-GCN~\cite{yan2018spatial} predicts future hand motion from pose history alone.
The forecasting models were trained using recorded task trajectories split into $288$ training trajectories, $17$ validation trajectories, and $19$ held-out test trajectories. 
We report \textit{precision}, \textit{recall}, and \textit{F1-score}, where a correct prediction corresponds to correctly anticipating the block that will be placed at the center. 

The linear extrapolation baseline achieved the lowest overall performance, indicating that simple kinematic heuristics are insufficient for reliable anticipation of upcoming placements. The immediate-placement baseline performed substantially better, showing that the most recently touched object is indeed informative about the likely next action. However, its performance remained below our full scene-aware model, indicating that anticipation cannot be reduced to touch detection alone.
The vanilla ST-GCN improved substantially over linear extrapolation, demonstrating that learned motion forecasting captures structure beyond simple kinematic heuristics; however, it did not surpass the immediate-placement baseline. In contrast, our scene-aware model achieved the best overall results, with the highest recall, precision, and F1-score among all compared approaches. These results show that conditioning motion forecasts on scene context substantially improves task-aligned anticipation of placement outcomes.


\begin{table}[t]
\centering
\caption{\textbf{Ablating next block placement forecasting performance.} 
Higher values indicate better results and more reliable support for predictive intervention.}
\label{tab:bench}
\begin{tabular}{lccc}
\hline
                              & \textbf{Recall (\%)} & \textbf{Precision (\%)} & \textbf{F1 (\%)} \\ \hline
\textbf{Linear Extrapolation} & 54.1                 & 71.7                    & 61.7             \\
\textbf{Instant Placement}    & 78.7                 & 96.0                    & 86.5             \\
\textbf{Vanilla ST-GCN}       & 65.6                 & 87.9                    & 75.1             \\
\textbf{Scene-aware ST-GCN}   & \textbf{86.9}        & \textbf{99.1}           & \textbf{92.6}    \\ \hline
\end{tabular}
\end{table}


\section{Discussion}

\subsection{Effectiveness of Predictive Feedback for Error Prevention}

Our results show that predictive feedback improves performance by intervening before errors are made. Compared to reactive feedback, it increased the success rate from $15\%$ to $65\%$, reduced edit distance from $2.06$ to $0.74$, and improved efficiency from $7.7$ to $8.7$ correctly placed blocks per minute. These gains were observed without increasing feedback frequency.
Conceptually, this shows the benefits of shifting feedback from a corrective to a preventive role. Reactive feedback is issued only after an incorrect placement has already occurred, forcing users to undo and recover from the mistake. Predictive feedback instead intervenes during the ongoing action, allowing users to redirect their motion before making the error. 
Importantly, the benefit comes from feedback timing rather than issuing more alerts, which is crucial as frequent, poorly timed feedback can disrupt task flow and reduce user acceptance~\cite{adamczyk2004if}. 
By reallocating feedback earlier in the action sequence, predictive feedback improves accuracy and efficiency while preserving the overall amount of system intervention.

\subsection{User Adaptation under Predictive Feedback}

Although predictive feedback significantly improved accuracy and efficiency, it did not reduce total task completion time. We observed that participants adapted their behavior in response to anticipatory interventions, becoming more cautious and sometimes pausing briefly to reassess their actions when warned. In this sense, predictive feedback appears to have shifted behavior away from rapid trial-and-error and toward more deliberate action.
This highlights an important property of predictive systems: users do not simply react to feedback, but incorporate it into their action planning. Thus, in our study, predictive feedback influenced not only task outcomes but also how participants approached the task, favoring error avoidance over speed.

\subsection{Complementary Roles of Feedback Modalities}

Although feedback modality did not significantly affect task performance, participants’ preferences suggest that different modalities supported the task in different ways. Audio feedback was often described as less disruptive because it allowed participants to keep visual attention on the workspace, whereas visual feedback provided clearer contextual information about what went wrong. Consistent with this interpretation, the combined audio+visual condition was most frequently preferred overall.
At the same time, these observations should be interpreted in the context of our task. Because the assembly pieces were primarily differentiated by color, auditory cues may have been especially effective as rapid warnings, while visual feedback mainly supported confirmation and interpretation. This makes modality a secondary design dimension relative to intervention timing, but still an important factor for usability and user preference.

\subsection{Design Trade-offs in Feedback Timing and Error Definition}

An important design choice in this study was the definition of error used to trigger feedback. Errors were defined at the level of completed placements on the assembly target, rather than at intermediate actions such as reaching toward or touching an incorrect block. This definition was intended to preserve exploratory behavior and user agency, allowing participants to handle, rearrange, or reconsider pieces without immediately triggering feedback. 

Alternative definitions, such as treating approaching an incorrect piece as errors, could enable even earlier intervention but would likely increase false positives and feedback frequency and reduce tolerance for exploration. In manual assembly tasks, such exploratory actions are often part of normal problem-solving behavior. Prior work on epistemic actions shows that people frequently manipulate objects to simplify cognition rather than to advance task completion, and that interrupting such actions can be counterproductive~\cite{kirsh1994distinguishing}. Treating these behaviors as errors may therefore lead to unnecessary interruptions and reduced user acceptance.

Our results suggest that combining a late error definition with predictive feedback provides a useful balance. Predictive alerts were issued early enough to prevent incorrect placements, while still allowing users flexibility in how they interacted with the pieces. This balance likely contributed to the system’s ability to improve accuracy without increasing feedback volume.
These findings highlight a broader design trade-off in anticipatory feedback systems: increasing sensitivity can enable earlier intervention but risks over-interruption, while more conservative definitions may better support user autonomy at the cost of delayed correction. Selecting an appropriate error definition should therefore depend on task demands, error costs, and the degree of exploration expected during interaction.
 
\subsection{Reliable Anticipation as a Requirement for Predictive Feedback}

Predictive feedback requires reliable task-aligned anticipation beyond simple heuristics. While the immediate-placement baseline confirms that recently touched objects are informative, its performance remained below that of the full scene-aware model. Likewise, linear extrapolation was insufficient for anticipating upcoming placements. These findings indicate that effective pre-commitment intervention requires more than detecting contact or extrapolating recent motion: the system must model how ongoing hand movement relates to the spatial configuration of task objects.
Our scene-aware forecaster achieved the strongest overall anticipation performance, showing that incorporating scene context is critical for deciding whether and when to intervene. 

\section{Limitations}

This study was conducted in a controlled table-top environment using a sequential assembly task. While this setting allowed precise measurement of error prevention and feedback timing, it does not capture the full complexity of real-world assembly scenarios involving varied object geometries, force interactions, or multi-step tool use. By focusing on a task in which motion cues alone are sufficient to understand user intent, our system is capable of accurately anticipating actions and reducing the amount of mistimed feedback. As more complex activities may require higher-level representations of user intent, such as action recognition or task-state modeling, the generalizability of our findings to more complex assembly tasks should be interpreted with caution. Nevertheless, in cases where user intent can be modeled accurately, we expect the benefits of predictive feedback to increase even further when faced with complex tasks that involve high error costs.

The study also involved a relatively small sample size (N=20) and short task durations, which are typical for controlled laboratory studies but limit the ability to assess longer-term learning effects or changes in user behavior over repeated sessions. Additionally, participants interacted with the system in a single session, and adaptation over extended use was not examined. However, we mitigated these effects by shuffling the order in which users encounter the different feedback modes, ensuring that, on average, they all occur at the same stage of a user's learning process. Finally, the system was evaluated in a fixed hardware configuration with an overhead camera and a predefined feedback setup. Performance and user experience may differ under alternative sensing configurations, display placements, or environmental conditions.

\section{Conclusion}

We presented \textsc{TRAFA}, the first real-time predictive feedback system for procedural tasks based on a \textit{Track-Forecast-Act} framework. By combining scene tracking, short-horizon hand motion forecasting, and intervention logic, the system can issue feedback before an incorrect placement is completed.
Across both technical benchmarking and a controlled user study, our results show that predictive feedback is both feasible and effective. The scene-aware forecasting model achieved the strongest task-aligned anticipation performance, and the interactive evaluation showed that predictive feedback significantly improved success rate, edit distance, and efficiency relative to reactive feedback and no feedback, without increasing feedback frequency.

The results show how real-time anticipation can be operationalized in an interactive system to shift feedback from post-hoc correction toward pre-commitment intervention. In this sense, we view the present system as a building block for more complex assistance problems: it isolates the core loop of tracking, forecasting, and intervening, and demonstrates that this loop can improve performance in a real interactive setting. This work is the first step toward predictive assistance systems that will inspire future work for richer procedural domains, including settings that require stronger task representations, more complex notions of user intent, and adaptive feedback policies that evolve with user behavior over time.




\bibliographystyle{ACM-Reference-Format}
\bibliography{reference}

\appendix

\section{Per-Participant Paired Differences}
\label{app:1}
\begin{figure*}[t]
  \centering
  \includegraphics[width=0.95\textwidth]{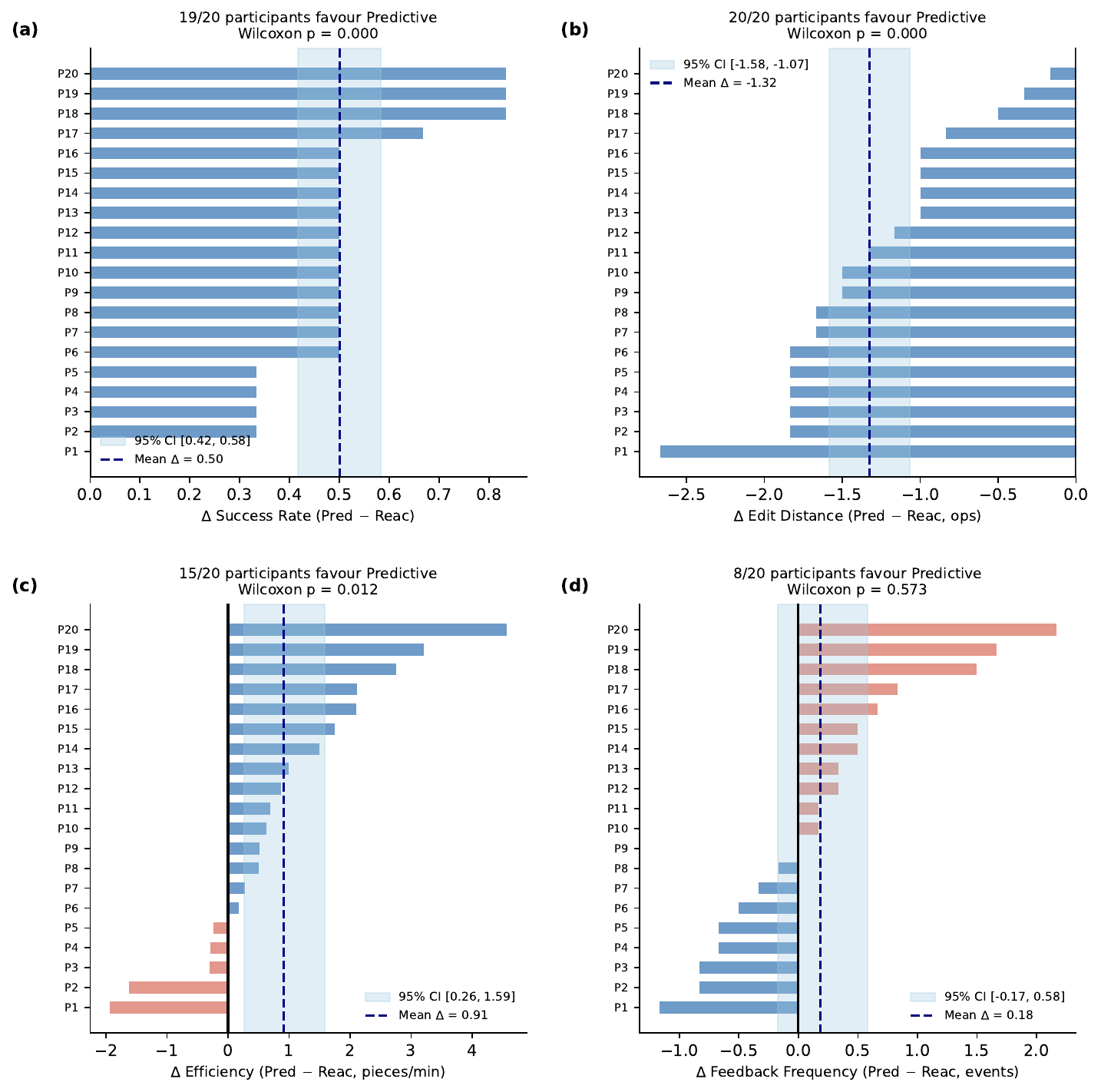}
 \caption{\textbf{Per-participant differences (Predictive-Reactive, aggregated) for four performance metrics (N=20).} Each bar represents one participant; blue bars indicate the participant favored the Predictive condition, red bars indicate the participant favored the Reactive condition. The shaded region shows the bootstrapped $95\%$ CI of the mean difference; the dashed line marks the mean $\Delta$. \textbf{(a)} Success Rate, \textbf{(b)} Edit Distance, \textbf{(c)} Efficiency, \textbf{(d)} Feedback Frequency. Statistical significance was assessed using two-sided Wilcoxon signed-rank tests.}
 \label{fig:paired_diff}
\end{figure*}

Figure~\ref{fig:paired_diff} shows the participant-level paired differences between the aggregated Predictive and Reactive conditions for the four primary metrics reported in Table~\ref{tab:res_sum}. The plots make the within-subject structure of the comparison explicit and are consistent with the patterns reported in Section~\ref{sec:eval}. For Success Rate (a) and Edit Distance (b), the effect is large and highly consistent across participants: 19/20 participants showed higher success under Predictive feedback, and 20/20 showed lower edit distance under Predictive feedback (both two-sided Wilcoxon signed-rank tests, $p<0.001$). For Efficiency (c), most participants (15/20) also favored Predictive feedback ($p=0.012$), although the smaller and more variable paired differences indicate greater between-participant variation in task pacing. In contrast, Feedback Frequency (d) did not differ significantly between conditions ($p=0.573$), indicating that Predictive and Reactive feedback triggered a comparable number of interventions. This supports the interpretation that the performance gains of Predictive feedback arise from improved timing of intervention rather than from issuing more feedback events.

\end{document}